\documentclass[a4paper,11pt]{article}\pdfoutput=1

\usepackage{amssymb,mathrsfs,amsmath}
\usepackage{a4wide}
\usepackage{color}
\usepackage[dvipsnames]{xcolor}
\usepackage{slashed,soul}
\usepackage{graphicx}
\usepackage{amsfonts}
\usepackage{lscape}
\usepackage[normalem]{ulem}
\def\linkcolor{cyan!70!black}

\usepackage[
colorlinks=true
,urlcolor=\linkcolor
,anchorcolor=\linkcolor
,citecolor=\linkcolor
,filecolor=\linkcolor
,linkcolor=\linkcolor
,menucolor=\linkcolor
,linktocpage=true
,pdfproducer=medialab
,pdfa=true
]{hyperref}
\usepackage{amsthm}
\usepackage{booktabs}
\usepackage{array}
\usepackage{rotating}
\usepackage[numbers, sort&compress]{natbib}
\usepackage{multirow}
\usepackage{float}
\usepackage[utf8]{inputenc}
\usepackage[T1]{fontenc}
\usepackage{appendix}
\usepackage{epsfig}
\usepackage{dsfont}
\usepackage{subfig}

\usepackage{soul,xcolor}
\setstcolor{red}

\newcommand{\be}{\begin{equation}}
\newcommand{\ee}{\end{equation}}

\newcommand{\beq}{\begin{equation}} 
\newcommand{\eeq}{\end{equation}} 
\newcommand{\ba}{\begin{array}}  
\newcommand{\ea}{\end{array}} 
\newcommand{\bea}{\begin{eqnarray}}  
\newcommand{\eea}{\end{eqnarray} }  
\newcommand{\bal}{\begin{align}}
\newcommand{\eal}{\end{align}}   
\newcommand{\bi}{\begin{itemize}}  
\newcommand{\ei}{\end{itemize}}  
\newcommand{\ben}{\begin{enumerate}}  
\newcommand{\een}{\end{enumerate}}  
\newcommand{\bc}{\begin{center}}
\newcommand{\ec}{\end{center}} 
\newcommand{\bt}{\begin{table}}
\newcommand{\et}{\end{table}}  
\newcommand{\btb}{\begin{tabular}}
\newcommand{\etb}{\end{tabular}}

\renewcommand{\baselinestretch}{1.2}

\allowdisplaybreaks

\let\OLDthebibliography\thebibliography
\renewcommand\thebibliography[1]{
  \OLDthebibliography{#1}
  \setlength{\parskip}{0pt}
  \setlength{\itemsep}{0pt plus 0.3ex}
}

\usepackage{fontawesome5}
\makeatletter
\newcommand{\github}[1]{%
   \href{#1}{\faGithubSquare}%
}
\makeatother

\begin{document}


\begin{titlepage}

\vspace*{-3.5truecm}
\begin{flushright}
IFT-UAM/CSIC-22-134
 \end{flushright}
\vspace{0.2truecm}

\begin{center}
\renewcommand{\baselinestretch}{1.8}\normalsize
\boldmath
{\LARGE\textbf{
Machine-Learned Exclusion Limits without Binning
}}
\unboldmath
\end{center}

\vspace{0.4truecm}

\renewcommand*{\thefootnote}{\fnsymbol{footnote}}

\begin{center}

{\bf 
Ernesto Arganda$\,\,^{a,b,c}$\footnote{\href{mailto:ernesto.arganda@uam.es}{ernesto.arganda@uam.es}}
Andres D. Perez,$\,^{a,b,c}$\footnote{\href{mailto:andresd.perez@uam.es}{andresd.perez@uam.es}}
Mart\'in de los Rios$\,^{a,b}$\footnote{\href{mailto:martin.delosrios@uam.es}{martin.delosrios@uam.es}}
\\and Rosa Mar\'{\i}a Sand\'a Seoane$\,^{b}$\footnote{\href{mailto:r.sanda@csic.es}{r.sanda@csic.es}}}

\vspace{0.5truecm}

{\footnotesize

$^a${\sl Departamento de Física Teórica, Universidad Autónoma de Madrid,\\ E-28049 Cantoblanco, Madrid, Spain\vspace{0.15truecm}}

$^b${\sl Instituto de Física Teórica UAM-CSIC, \\ C/ Nicol\'as Cabrera 13-15, Campus de Cantoblanco, 28049, Madrid, Spain \vspace{0.15truecm}}

$^c${\sl  IFLP, CONICET - Dpto. de Física, Universidad Nacional de La Plata, \\ C.C. 67, 1900 La Plata, Argentina \vspace{0.15truecm}}
}

\vspace*{2mm}

\end{center}

\renewcommand*{\thefootnote}{\arabic{footnote}}
\setcounter{footnote}{0}

\begin{abstract}
\noindent 

Machine-Learned Likelihoods (MLL) combines machine-learning classification techniques with likelihood-based inference tests to estimate the experimental sensitivity of high-dimensional data sets. We extend the MLL method by including Kernel Density Estimators (KDE) to avoid binning the classifier output to extract the resulting one-dimensional signal and background probability density functions. We first test our method on toy models generated with multivariate Gaussian distributions, where the true probability distribution functions are known. Later, we apply the method to two cases of interest at the LHC: a search for exotic Higgs bosons, and a $Z'$ boson decaying into lepton pairs. In contrast to physical-based quantities, the typical fluctuations of the ML outputs give non-smooth probability distributions for pure-signal and pure-background samples. The non-smoothness is propagated into
the density estimation due to the good performance and flexibility of the KDE method. We study its impact on the final significance computation, and we compare the results using the average of several independent ML output realizations, which allows us to obtain smoother distributions. We conclude that the significance estimation turns out to be not sensible to this issue.
\github{https://github.com/AndresDanielPerez/2211.04806-ML-Likelihood-with-KDE} 

\end{abstract}

\end{titlepage}

\tableofcontents

\section{Introduction}\label{sec:intro}

Modern machine learning (ML) has become a fundamental tool in experimental and phenomenological analyses of high-energy physics (for reviews see, for instance,~\cite{Larkoski:2017jix,Guest:2018yhq,Albertsson:2018maf,Radovic:2018dip,Carleo:2019ptp,Bourilkov:2019yoi,Karagiorgi:2021ngt,Feickert:2021ajf,Plehn:2022ftl} and for pioneer papers see~\cite{Denby:1987rk,Lonnblad:1990bi,Baldi:2014kfa}). The ML algorithms can be applied not only to event-by-event collider analyses but also used at the event-ensemble level~\cite{Metodiev:2017vrx,Khosa:2019kxd,Mullin:2019mmh,Chang:2020rtc,Flesher:2020kuy,Lai:2020byl,Nachman:2021yvi,Arganda:2021azw}. In order to estimate the experimental sensitivity to potential new-physics signals at colliders, several studies have recently appeared that combine the use of ML classifiers with traditional statistical tests~\cite{Cranmer:2015bka, Elwood:2018qsr,DAgnolo:2018cun,Nachman:2019dol,DAgnolo:2019vbw,Hollingsworth:2020kjg,Matchev:2020wwx,Cornell:2021gut,Aguilar-Saavedra:2021utu,dAgnolo:2021aun,Mikuni:2021nwn,Khosa:2022vxb,Letizia:2022xbe, Finke:2022lsu, Freitas:2022cno,Arganda:2022qzy}.
More specifically, it was shown in~\cite{Cranmer:2015bka} that the calibration of classifiers trained to distinguish signal and background samples under the relevant hypotheses ensures to proper estimate the likelihood ratio and consequently can be used to compute a statistical significance.

Recently, a simplification of~\cite{Cranmer:2015bka} has been proposed in~\cite{Arganda:2022qzy}, the so-called Machine-Learned Likelihoods (MLL), which computes the expected experimental sensitivity through the use of ML classifiers, utilizing the entire discriminant output.
A single ML classifier estimates the individual probability densities and subsequently one can calculate the statistical significance for a given number of signal and background events ($S$ and $B$, respectively) with traditional hypothesis tests. By construction, the output of the classifier is always one-dimensional, so we reduce the hypothesis test to a single parameter of interest, the signal strength $\mu$. On the one hand, it is simply and reliably applicable to any high-dimensional problem. On the other hand, using all the information available from the ML classifier does not require defining working points like traditional cut-based analyses.
The ATLAS and CMS Collaborations incorporate similar methods in their experimental analyses but consider only the classifier output as a good variable to bin and fit the binned likelihood formula (see, for instance, Refs.~\cite{ATLAS:2012byx,ATLAS:2015eiz,CMS:2017kxn,CMS:2018fdh,CMS:2018amb,CMS:2020bfa,ATLAS:2021kqb,ATLAS:2022ooq}).

The MLL code~\cite{MLLdiscovery-code} developed in~\cite{Arganda:2022qzy} only includes the calculation of the discovery hypothesis test, although the expressions needed to calculate the exclusion limits were provided. In~\cite{Arganda:2022mrd} we extend the MLL method by adding the exclusion hypothesis test. It is well-known that unbinned methods could provide a better performance than binned ones since the loss of information is minimized. In that sense, in this work we improve the MLL method with the use of Kernel Density Estimators (KDE)~\cite{RosenblattKDE,ParzenKDE}, in order to avoid binning the ML classifier output for extracting the resulting one-dimensional signal and background probability density functions (PDFs), as proposed in~\cite{Arganda:2022qzy, Arganda:2022mrd}. The implementation of unbinned methods to the ML output space has intrinsic difficulties that are usually not present if one considers physical based features, specifically the stochasticity of the machine learning training introduces fluctuations, even when the classifier approaches its optimal limit. These fluctuations translate to non-smooth distribution functions, that in turn, are propagated by the KDE into the density estimation given the plasticity of this consistent non-parametric method~\cite{Papamakarios:2019ccu}. Therefore, it is necessary to analyze the impact of the lack of smoothness in the statistical analysis. We propose to tackle this issue by working with a variable build from the average of several independent machine-learning realizations, that gives smoother PDFs.

We would like to highlight that binned methods are commonly used since one can usually optimize the binning to extract nearly all of the benefits of the unbinned approach, but this optimization can be a highly non-trivial and scenario-dependent task. The incorporation of KDE within our framework allows to automatically elude any binning optimization and outperform some of the most common binning schemes. For illustration, we compare the results of our unbinned MLL method with the results obtained by doing linear and non-linear binnings in the toy examples used to validate our setup, where the true PDFs are known.

The structure of the paper is the following: Section~\ref{sec:method} is devoted to summarizing the main features of the MLL method with the relevant expressions for the calculation of exclusion limits and the implementation of KDE in it. In Section~\ref{sec:apps} we show the performance of the MLL method with KDE and analyze the application of this unbinned method to the ML output space in different examples: in Section~\ref{sec:pdf} a case where the true probability density functions (PDFs) are known, through a toy model generated with multivariate Gaussian distributions; in Section~\ref{sec:heavyhiggs} we present an LHC analysis for the search for new heavy neutral Higgs bosons at $\sqrt{s}$ = 8 TeV and luminosity of 20 fb$^{-1}$, estimating not only exclusion limits, but also comparing our results with those report in~\cite{Baldi:2014kfa}; and in Section~\ref{sec:zprime} we present an HL-LHC study for Sequential Standard Model (SSM)~\cite{Altarelli:1989ff} $Z^\prime$ bosons decaying into lepton pairs, comparing the MLL+KDE performance for estimating 95\% CL exclusion limits with the results obtained applying a binned likelihood to the machine learning classifier output and also with respect to the projections reported by the ATLAS Collaboration for an LHC center-of-mass energy of $\sqrt{s} =$ 14 TeV with a total integrated luminosity of ${\cal L} =$ 3 ab$^{-1}$~\cite{ATLAS:2018tvr}. Finally, Section~\ref{sec:concl} summarizes our more important results and conclusions.

\section{Method}\label{sec:method}

In this section, we present the corresponding formulae for the estimation of exclusion sensitivities with the MLL method, first introduced in~\cite{Arganda:2022qzy, Arganda:2022mrd}. We summarize the main features of the method which allows dealing with data of arbitrarily high dimension through a simple ML classifier while using the traditional inference tests to compare a null hypothesis (the signal-plus-background one) against an alternative one (the background-only one). We also present the details of the implementation of KDE to obtain the unbinned posterior probability distributions from the classifier output, needed to compute the corresponding likelihood functions.

Following the statistical model in~\cite{Cranmer:2021urp}, we can define the likelihood $\mathcal{L}$ of $N$ independent measurements with an arbitrarily high-dimensional set of observables $x$ as
\begin{equation}
    \mathcal{L}(\mu,s,b) = \text{Poiss}\big(N|\mu S + B\big)\,\prod_{i=1}^{N}p(x_{i}|\mu,s,b)\label{eq:stat_model} \,,
\end{equation}
where $S$ ($B$) is the expected total signal (background) yield, Poiss stands for a Poisson probability mass function, and $p(x|\mu,s,b)$ is the probability density for a single measurement $x$, where $\mu$ defines the hypothesis we are testing for. 

We can model the probability density containing the event-by-event information as a mixture of signal and background densities
\begin{equation}
    p(x|\mu,s,b) = \frac{B}{\mu S + B}\,p_{b}(x)+\frac{\mu S}{\mu S + B}\,p_{s}(x)\,,\label{eq:prob_single_measurement}
\end{equation}
where $p_{s}(x)=p(x|s)$ and $p_{b}(x)=p(x|b)$ are, respectively, the signal and background probability density functions (PDFs) for a single measurement $x$, and $\frac{\mu S}{\mu S + B}$ and $\frac{B}{\mu S + B}$ are the probabilities of an event being sampled from the corresponding probability distributions.

To derive upper limits on $\mu$, and in particular considering additive new physics scenarios ($\mu\geq0$), we need to consider the following test statistic for exclusion limits~\cite{Cowan:2010js}:
\begin{equation}
\tilde{q}_{\mu} =   \begin{cases}
    0       & \quad \text{if } \hat{\mu} > \mu\,,\\
    -2\text{ Ln }\frac{\mathcal{L}(\mu,s,b)}{\mathcal{L}(\hat{\mu},s,b)}       & \quad \text{if } 0 \leq \hat{\mu} \leq \mu \,,\\
    -2\text{ Ln }\frac{\mathcal{L}(\mu,s,b)}{\mathcal{L}(0,s,b)}  & \quad \text{if } \hat{\mu} < 0\,,
  \end{cases}
  \label{eq:testexclusion1}
\end{equation}
where $\hat{\mu}$ is the parameter that maximizes the likelihood in Eq.~\eqref{eq:stat_model} 
\begin{equation}
    \sum_{i=1}^{N}\frac{p_{s}(x_{i})}{\hat{\mu}S\, p_{s}(x_{i}) +B\, p_{b}(x_{i})} = 1 \,.\label{eq:muhat_0}
\end{equation}
Considering our choice for the statistical model in Eq.~\eqref{eq:stat_model}, $\tilde{q}_{\mu}$ turns out 

\begin{equation}
\tilde{q}_{\mu} =   \begin{cases}
0 & \rm{if} \, \, \hat{\mu} > \mu \\
2(\mu-\hat{\mu}) S - 2 \sum_{i=1}^{N} \text{ Ln } \left( \frac{B p_b(x_i)+\mu S p_s(x_i)}{B p_b(x_i)+\hat{\mu} S p_s(x_i)}\right)  &  \rm{if} \, \, 0 \leq \hat{\mu} \leq \mu \\
2\mu S - 2 \sum_{i=1}^{N} \text{ Ln } \left( 1 + \frac{\mu S p_s(x_i)}{B p_b(x_i)}\right)  &  \rm{if} \, \,  \hat{\mu} < 0;
  \end{cases}
\label{eq:testdexclusion1}  
\end{equation}
Since $p_{s,b}(x)$ are typically not known, the base idea of our method in~\cite{Arganda:2022qzy} is to replace these densities for the one-dimensional manifolds that can be obtained for signal and background from a machine-learning classifier. After training the classifier with a large and balanced data set of signal and background events, it can be obtained the classification score $o(x)$ that maximizes the binary cross-entropy (BCE) and thus approaches~\cite{Neyman:1933wgr,Cranmer:2015bka}
\begin{equation}
    o(x) = \frac{p_{s}(x)}{p_{s}(x)+p_{b}(x)}\,,
    \label{optimal-classifier}
\end{equation}
as the classifier approaches its optimal performance. The dimensionality reduction can be done by dealing with $o(x)$ instead of $x$, using
\begin{equation}
    p_{s}(x) \rightarrow \tilde{p}_{s}(o(x))\,, \hspace{1cm} \text{and} \hspace{1cm} p_{b}(x) \rightarrow \tilde{p}_{b}(o(x))\,,
    \label{reduction}
\end{equation}
where $\tilde{p}_{s,b}(o(x))$ are the distributions of $o(x)$ for signal and background, obtained by evaluating the classifier on a set of pure signal or background events, respectively. 
Notice that this allows us to approximate both signal and background distributions individually, retaining the full information contained in both densities, without introducing any working point. These distributions are one-dimensional, and therefore can always be easily handled and incorporated into the test statistic in Eq.~\eqref{eq:testdexclusion1}
\begin{equation}
\tilde{q}_{\mu} =   \begin{cases}
0 & \rm{if} \, \, \hat{\mu} > \mu \\
2(\mu-\hat{\mu}) S - 2 \sum_{i=1}^{N} \text{ Ln } \left( \frac{B \tilde{p}_{b}(o(x_{i}))+\mu S \tilde{p}_{s}(o(x_{i}))}{B \tilde{p}_{b}(o(x_{i}))+\hat{\mu} S \tilde{p}_{s}(o(x_{i}))}\right)  &  \rm{if} \, \, 0 \leq \hat{\mu} \leq \mu \\
2\mu S - 2 \sum_{i=1}^{N} \text{ Ln } \left( 1 + \frac{\mu S \tilde{p}_{s}(o(x_{i}))}{B \tilde{p}_{b}(o(x_{i}))}\right)  &  \rm{if} \, \,  \hat{\mu} < 0;  
  \end{cases}
\label{eq:testexclusion2}
\end{equation}
as well as into the condition on $\hat{\mu}$ from Eq.~\eqref{eq:muhat_0}
\begin{equation}
    \sum_{i=1}^{N}\frac{\tilde{p}_{s}(o(x_{i}))}{\hat{\mu}S\, \tilde{p}_{s}(o(x_{i})) +B\, \tilde{p}_{b}(o(x_{i}))} = 1 \,.\label{eq:muhat}
\end{equation}

The test statistic in Eq.~\eqref{eq:testexclusion2} is estimated through a finite data set of $N$ events and thus has a probability distribution conditioned on the true unknown signal strength $\mu'$. 
For a given hypothesis described by the $\mu'$ value, we can estimate numerically the $\tilde{q}_{\mu}$ distribution. When the true hypothesis is assumed to be the background-only one ($\mu'=0$), the median expected exclusion significance $\text{med }[Z_{\mu}| 0]$ is defined as
\begin{equation}
    \text{med }[Z_{\mu}|0] = \sqrt{\text{med }[\tilde{q}_{\mu}|0]}\,,
\end{equation}
where we estimate the $\tilde{q}_{\mu}$ distribution by generating a set of pseudo-experiments with background-only events. 
Then, to set upper limits to a certain confidence level, we select the lowest $\mu$ which achieves the required median expected significance.

It is worth remarking that the output of the machine learning classifier, for a given set of events, gives us a sample of the desired PDFs $\tilde{p}_{s,b}(o(x))$.
Hence, to apply Eq.~\eqref{eq:testexclusion2} we first need to extract the classifier posteriors. 
As these samples are one-dimensional, we can always compute binned PDFs, as was done in~\cite{Arganda:2022qzy}. Binning the output variable is a typical procedure when using ML tools. Nevertheless, it is also possible to compute the PDFs through other parametric (such as Mixture Models \citep{MixtureModels}) or non-parametric methods (such as Kernel Density Estimation (KDE)~\cite{RosenblattKDE,ParzenKDE} or Neural Density Estimation \citep{Papamakarios:2019ccu}).
In comparison with other density-estimation methods, KDE has the advantage of not
assuming any functional form for the PDF, in contrast with the mixture of Gaussian methods,
while keeping the computation and the interpretation simple, as opposed to neural density
estimation methods.
For this reason, in this work, we made extensive use of the KDE method\footnote{During the completion of this work, Ref.~\cite{GomezAmbrosio:2022mpm} has appeared, in which data-driven methods are used for dealing with unbinned multivariate EFT observables. Although this work is similar to ours in the spirit of avoiding the binning of multidimensional variables, there are significant differences in the application of ML and KDE algorithms between our MLL+KDE method and the ML4EFT framework of~\cite{GomezAmbrosio:2022mpm}.}, through its \texttt{scikit-learn} implementation \citep{scikit-learn}.

Given a set of $N$ events that were previously classified by the machine learning as signal (background) events, the PDF estimated by the KDE method is defined as
\begin{equation}
    p_{s,b}(o(x)) = \frac{1}{N}\sum_{i}^{N}\kappa_{\epsilon} \left[o(x) - o(x_{i}) \right]
\end{equation}
where $\kappa_{\epsilon}$ is a kernel function that depends on the "smoothing" scale, or bandwidth parameter $\epsilon$. There are several different options for the kernel function.
In this work, we used the \textit{Epanechnikov kernel} \citep{VAEpanechnikov} as it is known to be the most efficient kernel \citep{Papamakarios:2019ccu}.
This kernel is defined as

\[
    \kappa_{\epsilon}(u)= 
\begin{cases}
    \frac{1}{\epsilon}\frac{3}{4}\left(1 - (u/\epsilon)^{2} \right),& \text{if } |u|\leq \epsilon\\
    0,              & \text{otherwise}
\end{cases}
\]

It is important to remark that the "bandwidth" parameter $\epsilon$ censors the degree of smoothness. Hence, a very low $\epsilon$ will overfit the data, whereas a very high $\epsilon$ will underfit it. 
In all our examples the $\epsilon$ was selected through a grid search done using the \texttt{GridSearchCV} function inside the \texttt{sklearn.model\textunderscore selection}
python package. Given a value for $\epsilon$, this function estimates the log-likelihood of the
data using a 5-fold cross-validation strategy, i.e. the data set is split into 5 smaller sets, 4 are
used to fit the KDE which is then validated on the remaining part of the data. Finally, the
function gives as an output the $\epsilon$ which maximizes the data likelihood.
Also is worth remarking that although KDEs method suffers from the curse of dimensionality, we are applying such technique to the one-dimensional output of the machine learning classifier to avoid this problem.

Notice that the machine learning training (and hence the machine learning predictions) is a stochastic process that introduces small fluctuations around the optimal limit. These in turn could translate to non-smooth PDFs. To tackle this issue, the same procedure described above can be done when using an ensemble of $N$ base classifiers trained on random subsets of the original data set, that average their individual predictions to form a final prediction. In this case, $o(x)$ can simply be replaced by $<o(x)> = \frac{1}{N}\sum_{i}^{N}o_{i}(x)$, which in turns gives smoother PDFs $\tilde{p}_{s,b}(<o(x)>)$.

For completeness, we also introduce here the median exclusion significance estimation for the traditional Binned Likelihood (BL) method and the use of Asimov data sets~\cite{Cowan:2010js}, which will be used to compare our technique
\begin{equation}
    {\rm med}[Z_\mu|0]=\sqrt{\tilde{q}_{\mu}}=\left[2\sum_{d=1}^{D}\left(B_d\text{ Ln }\left(\frac{B_d}{S_d+B_d}\right)+S_d\right)\right]^{1/2}\,,
    \label{binned-Z}
\end{equation}
where $S_{d}$ and $B_{d}$ are the expected number of signal and background events in each bin $d$. This approximation is very effective but runs into trouble when the dimension of the data grows, which is known as the curse of dimensionality since the number of data points required to reliably populate the bins scales exponentially with the dimension of the observables $x$. 
This is a non-existent problem in our method, which always reduces the original dimension to one as stated in Eq.~\eqref{reduction}, allowing the application of the BL method to the classifier output, as also done by experimental collaborations when using ML methods, as mentioned in Section~\ref{sec:intro}.

\section{Application examples}\label{sec:apps}

\subsection{Known true PDFs: multivariate Gaussian distributions}\label{sec:pdf}

To show the performance of the MLL method with KDE we first analyze toy models generated with multivariate Gaussian distributions of different dimensions,

\begin{equation}
\mathcal{N}_{dim}(\boldsymbol m,\boldsymbol\Sigma) = \frac{1}{(2\pi)^{n/2} |\boldsymbol\Sigma|^{1/2}} exp\left( -\frac{1}{2}(x-m)^{T} \boldsymbol\Sigma^{-1} (x-m) \right),
\end{equation}
with mean $\boldsymbol m$, and covariance matrix $\boldsymbol\Sigma$.

\begin{figure}
    \begin{minipage}{0.5\linewidth}
        \centering
        \includegraphics[width=0.75\linewidth]{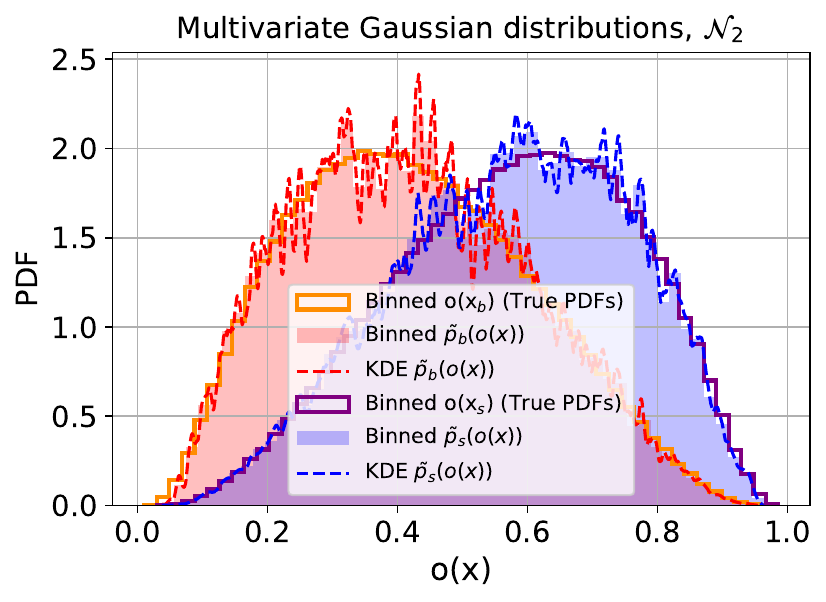}
         \includegraphics[width=0.75\linewidth]{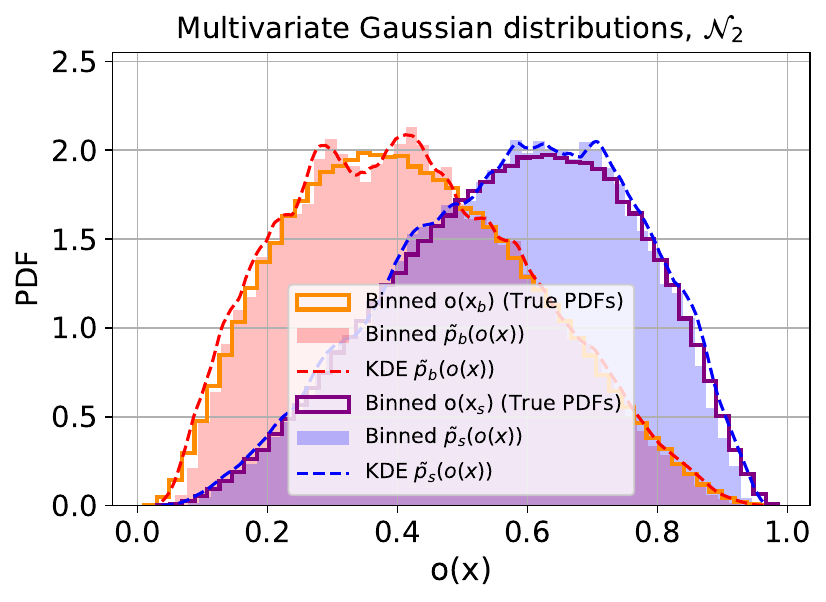}
    \end{minipage}%
    \hfill
    \begin{minipage}{0.5\linewidth}
        \centering
        \includegraphics[width=1\linewidth]{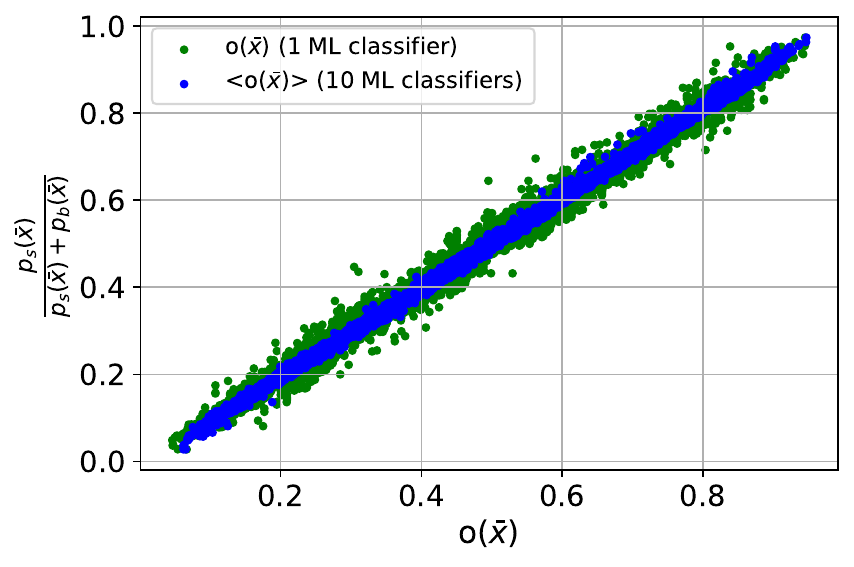}
    \end{minipage}%
\caption{Results for the $\mathcal{N}_{2}(\boldsymbol m,\boldsymbol\Sigma)$ case. \textit{Top left panel:} Output of a single {\tt XGBoost} classifier. \textit{Bottom left panel:} Averaged output of 10 {\tt XGBoost} classifiers, defined as $o(x) = \frac{1}{10}\sum_{i}^{10}o_{i}(x)$. \textit{Right panel:} Comparison between our trained classifier output and the mathematically optimal performance defined in Eq.~\eqref{optimal-classifier}.}

\label{fig:test}
\end{figure}

We start with the simplest case, consisting of an abstract two-dimensional space $(x_1,x_2)$. Events are generated by Gaussian distributions $\mathcal{N}_{2}(\boldsymbol m,\boldsymbol\Sigma)$, with $\boldsymbol m = +0.3 (-0.3)$ and no correlation, i.e., covariance matrices $\boldsymbol\Sigma = \mathbb{I}_{2\times 2}$ for $S$ ($B$). We trained supervised per-event classifiers, {\tt XGBoost}, with 1M events per class (balanced data set), to distinguish $S$ from $B$.
The PDFs obtained from the classifier output, $o(x)$, can be found in the top left panel of Figure~\ref{fig:test}, for two new independent data sets of pure signal (blue) and pure background (red) events. 

Since in this example we know the true underlying distributions in the original multidimensional space, we can test Eq.~\eqref{optimal-classifier}.
In the right panel of Figure~\ref{fig:test} we show, in green dots, the output of one machine learning realization vs. the right-hand-side of Eq.~\eqref{optimal-classifier} estimated with the real signal and background probability functions.
We can observe that the classifier approaches the optimal limit, although there are some small fluctuations around the 1-to-1 line. These fluctuations are independent of the sampling of the data and come from the stochasticity inherent to any machine learning training process. In turn, these fluctuations translate to non-smooth PDFs for the machine learning output of background and signal events, as can be seen in the red and blue shadow histograms in the top left panel of Figure~\ref{fig:test}.

As explained before, to solve this issue we can take advantage of ensembles, and build a variable from the average output of ten independent machine learning realizations, define as $<o(x)> = \frac{1}{10}\sum_{i}^{10}o_{i}(x)$. 
It can be seen in the red and blue shadow histograms of the bottom left panel of Figure~\ref{fig:test} that, with this definition, the small fluctuations are washed out resulting in smoother PDFs. For completeness on both left panels, we also present the estimations of $\tilde{p}_{s,b}(o(x))$ using the true PDFs (orange and purple solid lines), the KDE  over the machine learning output $o(x)$ (red and blue dashed curves of top left panel), and KDE over the average variable $<o(x)>$ (red and blue dashed lines of the bottom left panel). 
On the one hand, it can be seen that, due to the flexibility of the KDE method, when fitting the machine learning output $o(x)$ the resulting distributions follows the fluctuations around the true PDFs. On the other hand, it can be seen that, the KDE distributions obtained when fitting the average variable are smooth and closely approach the true PDFs.

\begin{figure}[t]
\centering
  \includegraphics[width=.48\linewidth]{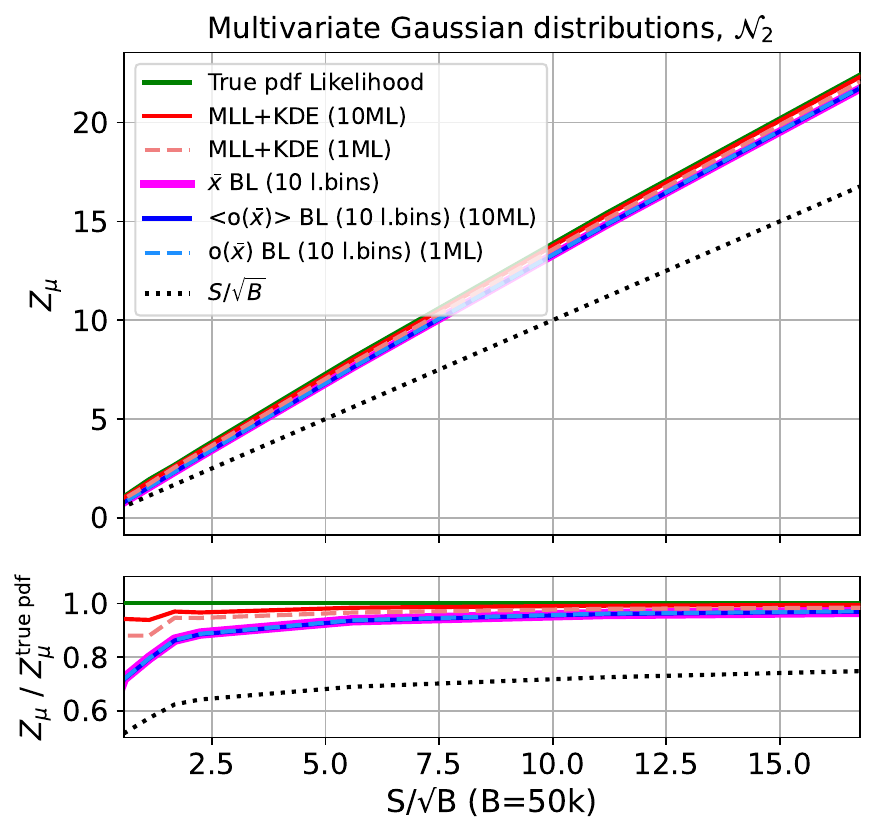}
\caption{Exclusion-limit significance for $\mathcal{N}_{2}(\boldsymbol m,\boldsymbol\Sigma)$ with $ \boldsymbol m= +0.3 (-0.3)$ for $S$ ($B$) and no correlation, for fixed $\langle B \rangle=50$k, and different signal strengths $\langle S \rangle$.  The red curves show the result of implementing the MLL+KDE method, while the blue and magenta curves represent the results obtained by applying the BL method to the classifier's one-dimensional output and the original two-dimensional space, respectively. Dashed curves use the output of a single classifier, while solid lines use the averaged output of 10 classifiers. For comparison, we include the green solid curve with the results obtained using the true PDFs.
}
\label{fig:X4}
\end{figure}

In Figure~\ref{fig:X4} we show the results for the MLL exclusion significance with KDE considering an example with a fixed background of $\langle B \rangle=50$k and different signal strengths. We also include the significance calculated using the true probability density functions in Eq.~\eqref{eq:testexclusion1}, and the results employing a binned Poisson log-likelihood of the original two-dimensional space $(x_1,x_2)$ with Eq.~\eqref{binned-Z}, which is possible to compute in this simple scenario. For completeness, we also include the results binning the one-dimensional ML output variable for obtaining the PDFs as in~\cite{Arganda:2022qzy, Arganda:2022mrd}. As can be seen, since we are analyzing a simple example, the significances estimated with all the methods are indistinguishable from the ones estimated with the true PDFs, which is expected given the low dimensionality of the space.

We would like to highlight that the significance does not change significantly if we employ either $o(x)$ computed with a single ML classifier, or with the averaged variable $<o(x)>$ calculated ensembling several ML trainings. In addition, for both the MLL+KDE and the true PDF methods, the significance is estimated by generating a set of pseudo-experiments with a finite-size number of events. This introduces a small statistical fluctuation due to the randomness of the sample.

\begin{figure}
  \centering
  \includegraphics[width=0.7\textwidth]{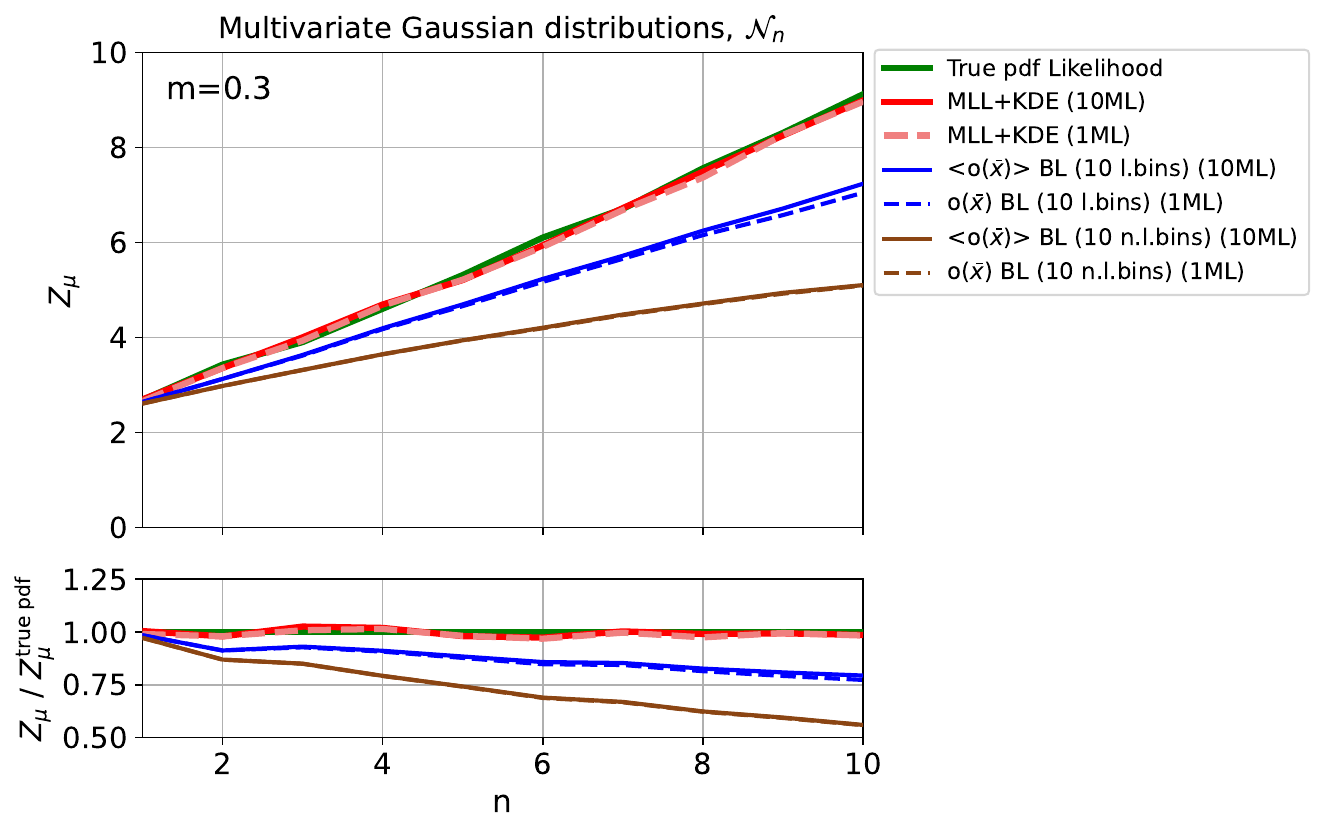}
\caption{Exclusion-limit significance for $\mathcal{N}_{dim}(\boldsymbol m,\boldsymbol\Sigma)$ with $ \boldsymbol m= +0.3 (-0.3)$ for $S$ ($B$) and no correlation, as a function of the $dim$, for fixed $\langle B \rangle=50$k and $\langle S \rangle=500$. The red curves show the result of implementing the MLL+KDE method, while the blue and brown curves represent the results obtained by applying the BL method to the classifier one-dimensional output for 10 linear and no-linear bins, respectively. Dashed curves use the output of a single classifier, while solid lines use the averaged output of 10 classifiers. For comparison, we include the green solid curve with the results obtained using the true PDFs.
}
\label{dimgaussian-Z}
\end{figure}

The advantage of the MLL+KDE method against traditional approaches appears when dealing with $dim=n$, with $n>2$. 
In Figure~\ref{dimgaussian-Z} we present the exclusion significance for higher dimensional data generated with $\mathcal{N}_{n}(\boldsymbol m,\boldsymbol\Sigma)$, no correlation $\boldsymbol\Sigma = \mathbb{I}_{n\times n}$, and $\boldsymbol m = +0.3 (-0.3)$ for $S$ ($B$).

It is worth reminding that the Binned-Poisson Likelihood method becomes intractable in the original high-dimensional space. 
Also, it is interesting to note that, the results with the MLL+KDE method approach the ones with the true generative functions for all the analyzed dimensions.
It is important to highlight that the ML output is always one-dimensional regardless of the dimension of the input data and, hence, can always be easily binned.
For completeness, we show in Figure~\ref{dimgaussian-Z} the significances obtained by applying a BL method to the machine learning output with two different types of binning: a linear binning where all bins have the same size (in the one-dimensional output space), and a standard non-linear approach where all bins have the same number of background events (a binning strategy typically used by experimental collaborations since it avoids the presence of low-statistic bins in the background estimation, which in turns constraints systematic uncertainties). As can be seen, binning the output of the machine learning results in a non-negligible drop in significance. This can be understood as the binning introduce a loss of information due to a resolution effect. For this example, the linear binning turns out to be more effective for the BL method. In addition, and as in the $n=2$ example, for MLL+KDE  the use of an ensemble of machine learning realizations to obtain smoother PDFs does not change the results obtained with one single classifier. The same is verified when using the BL method over the average variable, although this behavior is expected since this method creates histograms from the distributions.

\begin{figure}
  \centering
  \includegraphics[width=0.48\textwidth]{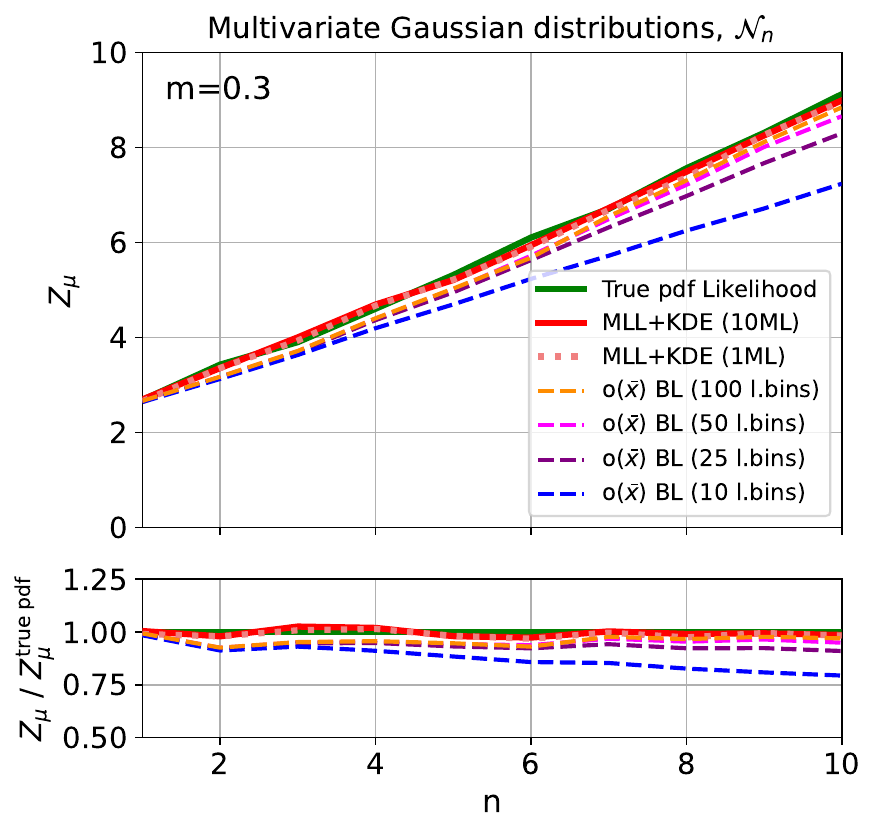}
  \includegraphics[width=0.48\textwidth]{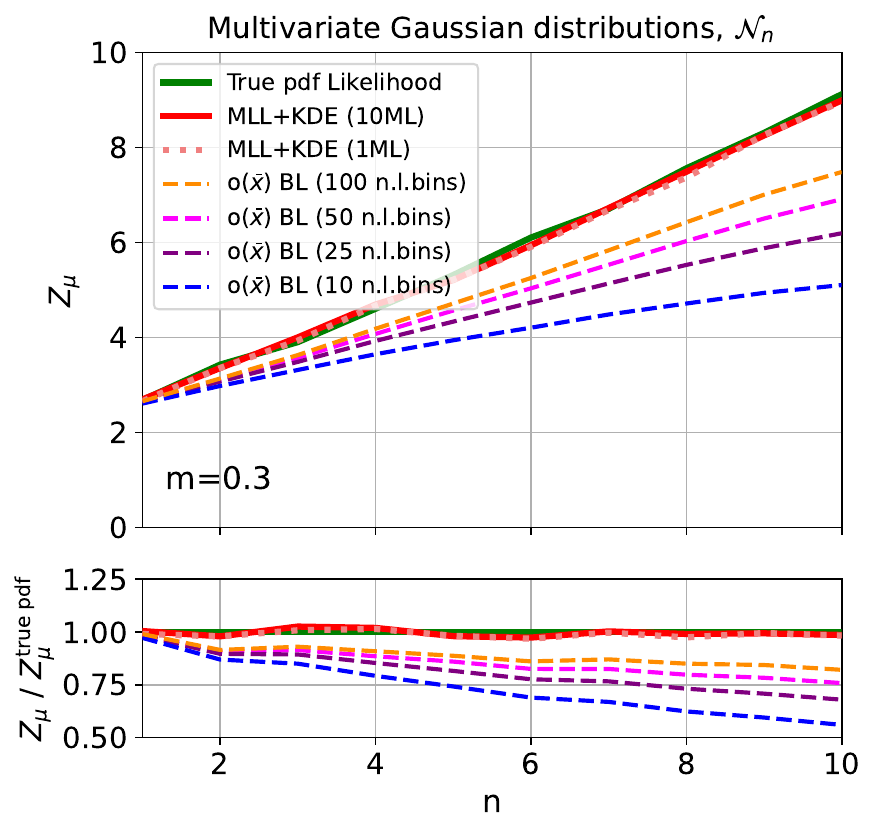}
\caption{Exclusion-limit significance for $\mathcal{N}_{dim}(\boldsymbol m,\boldsymbol\Sigma)$ with $ \boldsymbol m= +0.3 (-0.3)$ for $S$ ($B$) and no correlation, as a function of the $dim$, for fixed $\langle B \rangle=50$k and $\langle S \rangle=500$.  The red solid curve shows the result of implementing the MLL+KDE method, while the green curve shows the results obtained using the true PDFs.
Dashed color curves represent the results obtained by applying the BL method to the classifier's one-dimensional output for different bin numbers.
\textit{Left panel:} linear binning. 
\textit{Right panel:} non-linear binning (same number of B events per bin).}
\label{dimgaussian-Zbins}
\end{figure}

In the left and right panels of Figure~\ref{dimgaussian-Zbins} we show the impact in the previous example of increasing the number of bins when applying BL to the classifier output, both for linear and non-linear bins, respectively. As stated before, linear binning proves to be a better sampling choice since its result approaches the ones obtained with the MLL+KDE method and with the true PDFs, when increasing the number of bins. Regarding the bins with the same number of background events, even though performance improves with more bins, the results are worst than its linear binning counterpart. This example shows the difficulties arising when trying to find an optimal binning that is not known a priori, and this highlights the advantage of using MLL+KDE, which although computationally expensive (when tuning the bandwidth parameter), sets an upper limit in the significance that can be achieved. 
It is also possible to automatically choose the optimal number of bins for histograms, as well as to tune the width of each bin, in a similar fashion as done for the $\epsilon$ parameter in the KDE method. The results of this analysis can be found in Appendix~\ref{sec:appendix}, where we show that the significances obtained optimizing the bin widths is similar to the ones assuming equal-sized bins, and hence, the MLL-KDE method still offers the best significances when
compared to different binned multivariate approaches.

\begin{figure}
  \centering
  \includegraphics[width=0.48\textwidth]{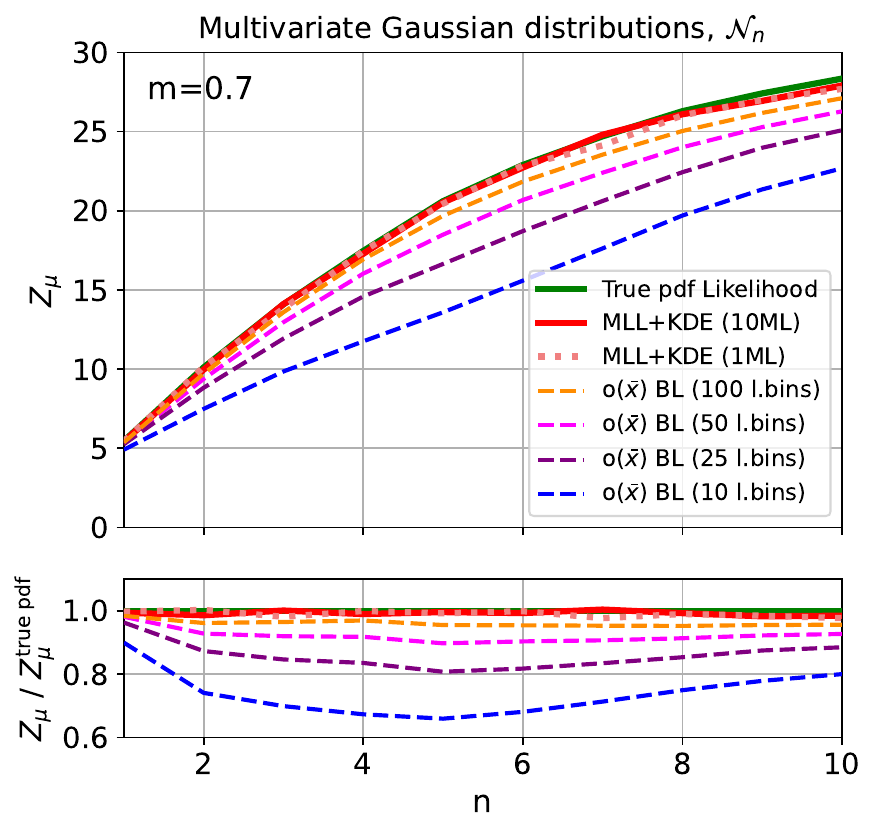}
  \includegraphics[width=0.48\textwidth]{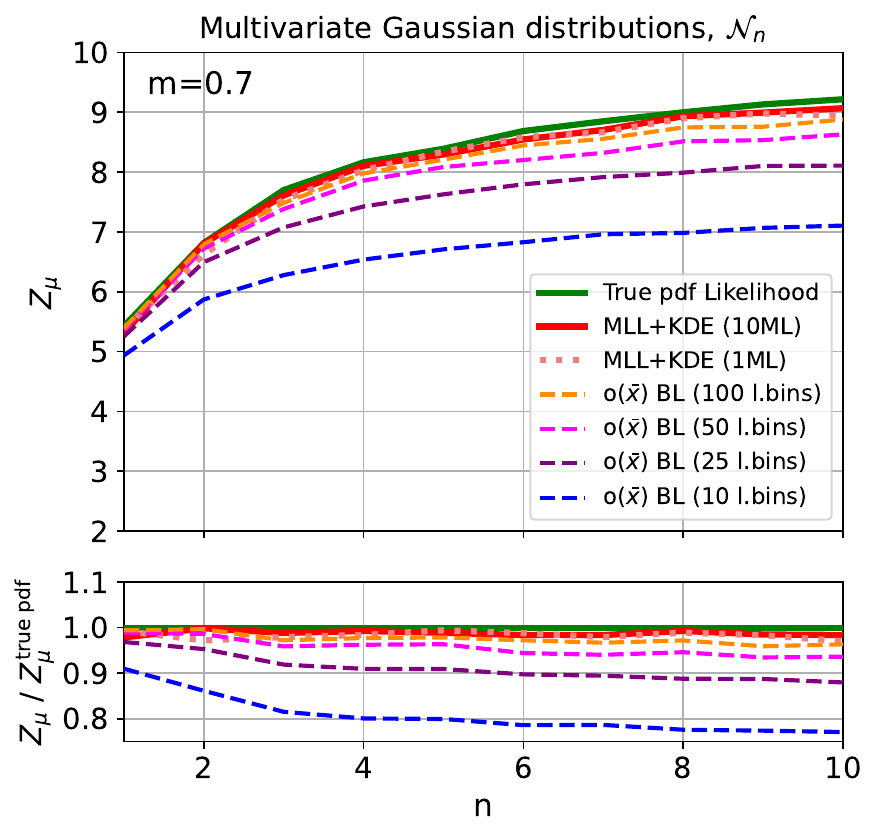}
\caption{Exclusion-limit significance for $\mathcal{N}_{dim}(\boldsymbol m,\boldsymbol\Sigma)$ with $ \boldsymbol m= +0.7 (-0.7)$ for $S$ ($B$), as a function of the $dim$, for fixed $\langle B \rangle=50$k and $\langle S \rangle=500$.  The red solid curve shows the result of implementing the MLL+KDE method, while the green curve shows the results obtained using the true PDFs.
Dashed color curves represent the results obtained by applying the BL method (with linear bins) to the classifier one-dimensional output for different bin numbers.
\textit{Left panel:} covariance matrices $\boldsymbol\Sigma = \mathbb{I}_{2\times 2}$ (no correlation). 
\textit{Right panel:} covariance matrices $\boldsymbol\Sigma_{ij}=1$ if $i=j$ and 0.5 if $i\neq j$.}
\label{dimgaussian-Z-cov}
\end{figure}

Finally, in the right panel of Figure~\ref{dimgaussian-Z-cov} we show a case with correlation,  $\mathcal{N}_{n}(\boldsymbol m,\boldsymbol\Sigma)$, with $\boldsymbol m = +0.7 (-0.7)$ for $S$ ($B$), and $\boldsymbol\Sigma_{ij}=1$ if $i=j$ and 0.5 if $i\neq j$. Comparing with the same example without correlation in the left panel of Figure~\ref{dimgaussian-Z-cov}, the correlation makes the signal and background more difficult to distinguish, hence we obtain lower significance values, with MLL+KDE still offering the best performance.

Although these are toy models they allow us to understand the performance of MLL with KDE method over problems of different complexity and demonstrate its improvement with respect to the BL method applied to the classifier output. Particularly the MLL+KDE has a stable behavior when increasing the dimensionality of the input space, as well as when increasing the separation of the signal and background distributions on the original abstract variables. On the other hand, the BL method applied to the classifier output departs from the results obtained with the true PDFs as the number of dimensions and separation of signal and background samples increases. The number of bins to use is another limitation, non-existent in our method that uses a non-parametric technique for PDF extraction. We also tested that although the KDE method is sensible to the fluctuations inherent to the machine learning classifier output, the lack of smoothness of the extracted PDFs does not affect the estimation of the significance within our framework.

\subsection{New exotic Higgs bosons at the LHC} \label{sec:heavyhiggs}

In this section, we apply our method in the search for an exotic electrically-neutral heavy Higgs boson ($H^{0}$) at the LHC, which subsequently decays to a $W$ boson and a heavy electrically-charged Higgs boson ($H^{\pm}$). This example was first analyzed with machine learning methods in Ref.~\cite{Baldi:2014kfa}. The exotic $H^{\pm}$ decays to another $W$ boson and the SM Higgs boson ($h$). Taking into account only the dominant decay of the Higgs boson, the signal process is defined as
\begin{equation}
    gg\to H^{0}\to W^{\mp}H^{\pm}\to W^{\mp}W^{\pm}h\to W^{\mp}W^{\pm}b\bar{b}.
    \label{signalheavyH}
\end{equation}
The background is therefore dominated by top-pair production, which also gives $W^{\mp}W^{\pm}b\bar{b}$.

For our analysis, we use the same data presented in~\cite{Baldi:2014kfa} that is publicly available at \cite{higgsdataset}, which focus on the semi-leptonic decay mode of both background and signal events (one $W$ boson decaying leptonically and the other one decaying hadronically), giving as final state $\ell\nu jj b\Bar{b}$. The data set consists of low-level variables (twenty-one in total, considering the momentum of each visible particle, the $b$-tagged information of all jets, and the reconstruction of the missing energy) and seven high-level variables ($m_{jj}$, $m_{jjj}$, $m_{\ell\nu}$, $m_{j\ell\nu}$, $m_{b\Bar{b}}$, $m_{Wb\Bar{b}}$ and $m_{WWb\Bar{b}}$), expected to have higher discrimination power between signal and background (see~\cite{Baldi:2014kfa} for more details). The signal benchmark case corresponds to a $m_{H^{0}}=425$ GeV and $m_{H^{\pm}}=325$ GeV.

\begin{figure}
  \centering
  \includegraphics[width=0.48\textwidth]{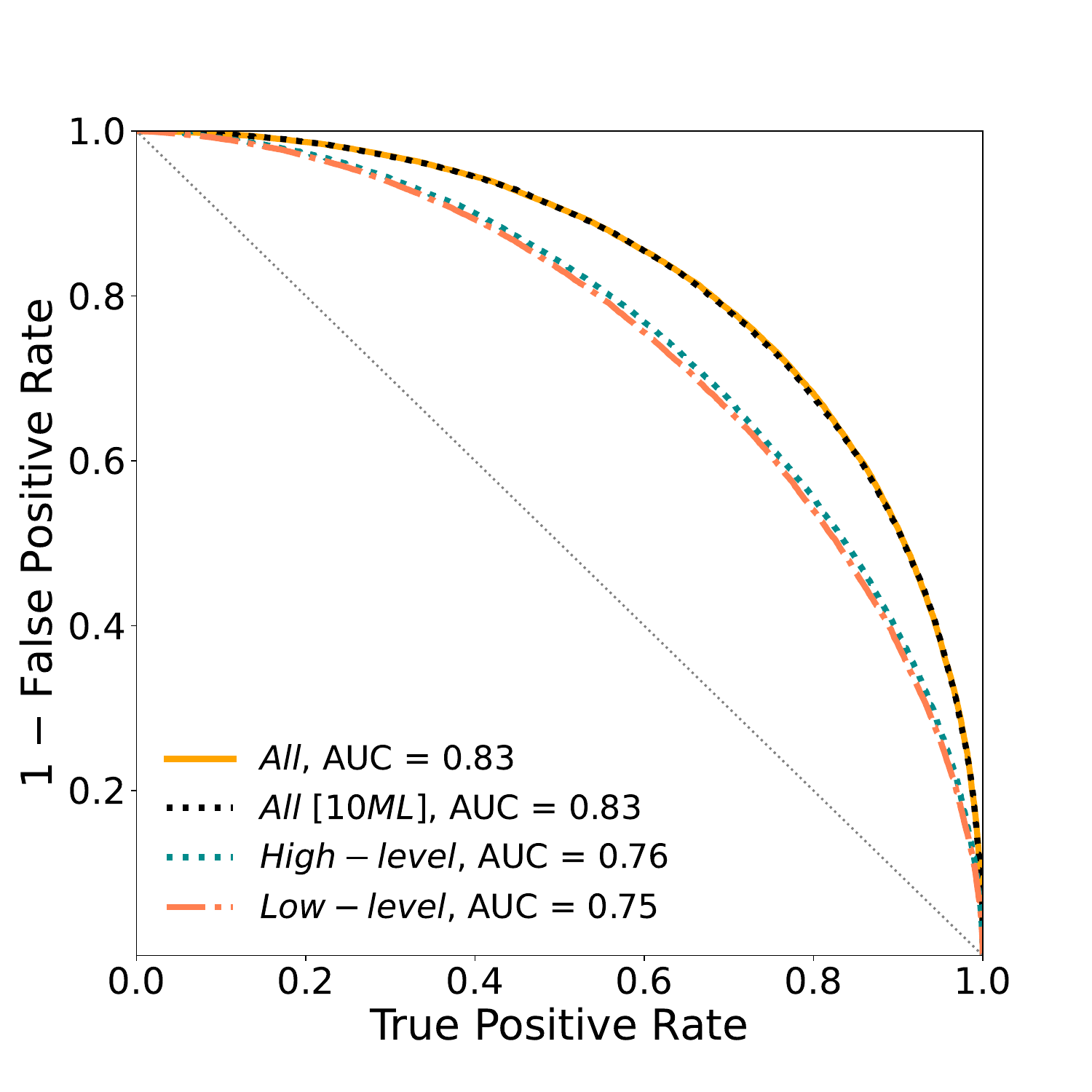}
  \includegraphics[width=0.48\textwidth]{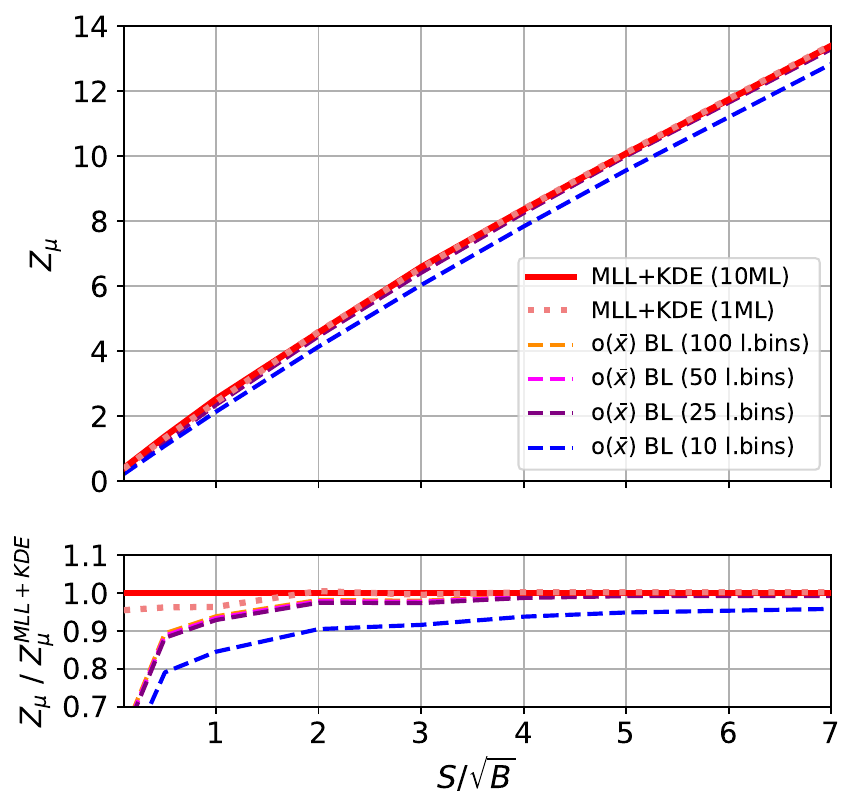}
\caption{\textit{Left panel:} ROC curves for the {\tt XGBoost} classifiers associated with each possible data representation, trained to discriminate between $H^{0}$ and $t\Bar{t}$ productions. \textit{Right panel:} Exclusion limits for the search for a heavy neutral $H^{0}$ with MLL+KDE method (red solid line corresponds to the averaged variable of 10 ML, and dotted orange line corresponds to $1$ ML), and with the BL  fit for different number of linear bins (dashed curves), for fixed $\langle B \rangle=86$k, and different signal strengths $\langle S \rangle$.}
\label{fig:HIGGS}
\end{figure}

For this example, we have trained three {\tt XGBoost} classifiers with three different data representations: only low-level variables, only high-level variables, and combining both low and high-level features. 
For completeness we also add the result obtained when using an average variable obtained after ensembling $10$ ML classifiers with all the input variables.
In the left panel of Figure~\ref{fig:HIGGS} we show the ROC curves for the analysis, and as expected, the best performance was achieved using both low and high-level features (for both the averaged and non-averaged variable). These results are in agreement with the analysis performed in~\cite{Baldi:2014kfa}, obtained with different ML algorithms. In the following, we will work with the latter data representation to estimate the expected significance for the search for heavy Higgs.

To compute the expected background yield at the ATLAS detector at $\sqrt{s}$=8 TeV and luminosity of 20 fb$^{-1}$, $B\simeq86$k, we simulated background events with {\tt MadGraph5\_aMC@NLO 2.6}~\cite{Alwall:2014hca},  using {\tt PYTHIA 8}~\cite{Sjostrand:2014zea} for showering and hadronization, and {\tt Delphes 3}~\cite{deFavereau:2013fsa} for fast detector simulation. We applied all the selection cuts in~\cite{Baldi:2014kfa}, and checked that the different kinematic distributions from our simulation are in agreement with the ones from the public data set. With the expected background prediction, we scan over the expected signal yield, $S$, to be agnostic regarding the coupling values of the model.

The exclusion significance for different $\frac{S}{\sqrt{B}}$ ratios are shown in the right panel of Figure~\ref{fig:HIGGS}. The results for the MLL+KDE methods do not yield significant differences and are shown as the red solid curve for the averaged variable of 10 ML and as dotted orange curve correspond for $1$ ML. We also present as dashed curves the significance binning the one-dimensional ML output for different numbers of bins. We would like to remark that binning the original feature space is not possible due to its high dimensionality (twenty-one and seven low and high-level variables, respectively). Also for this collider example, it can be seen that the MLL+KDE method outperforms the results obtained with the binned likelihood procedure.

\begin{table}
\centering
\begin{tabular}{ cc } 
 \hline
 Method & $\sigma_{\text{fid}}$(pb) ($95\%$ C.L. upper limit) \\ 
  \hline
 MLL+KDE & $8.94\times 10^{-3}$ \\ 
 MLL+KDE (10ML) & $8.84\times 10^{-3}$ \\ 
 o$(\bar{x})$ BL (100 bins) & $9.91\times 10^{-3}$ \\ 
 o$(\bar{x})$ BL (50 bins) & $9.97\times 10^{-3}$ \\ 
 o$(\bar{x})$ BL (25 bins) & $10.03\times 10^{-3}$ \\ 
 o$(\bar{x})$ BL (10 bins) & $11.15\times 10^{-3}$ \\ 
 \hline
\end{tabular}
 \caption{Expected cross-section upper limit at 95$\%$ C.L., considering ATLAS detector, $\sqrt{s}$=8 TeV and luminosity of 20 fb$^{-1}$ ($B=86$k). Background process and cuts as discussed in the main text.}
\label{table:HH-upperlimits}
\end{table}

Since no excess has been found, we can compute the expected cross-section upper limit at 95$\%$C.L. for the new exotic Higgs bosons search, which corresponds to the value of $\frac{S}{\sqrt{B}}$ that gives $Z=1.64$. The results are presented in Table~\ref{table:HH-upperlimits}.

\begin{table}
\centering
\begin{tabular}{ cc } 
 \hline
 Method & Z \\ 
  \hline
 NN~\cite{Baldi:2014kfa} & 3.7$\sigma$ \\ 
 DN~\cite{Baldi:2014kfa} & 5.0$\sigma$ \\ [0.2cm]
 MLL+KDE & 6.61$\sigma$ \\ 
 MLL+KDE (10ML) & 6.65$\sigma$ \\ 
 o$(\bar{x})$ BL (100 bins) & 6.53$\sigma$  \\ 
 o$(\bar{x})$ BL (50 bins) & 6.52$\sigma$ \\ 
 o$(\bar{x})$ BL (25 bins) & 6.43$\sigma$ \\ 
 o$(\bar{x})$ BL (10 bins) & 6.14$\sigma$ \\ 
 \hline
\end{tabular}
\caption{Discovery significances assuming $B=1000$ and $S=100$. For comparison, we also include the results for the same case shown in~\cite{Baldi:2014kfa} using a shallow neural network (NN) and a deep neural network (DN).}
\label{table:HH-Zcomparison}
\end{table}

For completeness, and to compare with the results of~\cite{Baldi:2014kfa}, we show in Table~\ref{table:HH-Zcomparison} the discovery significance for MLL+KDE and BL methods. Notice that for this calculation we artificially set $B=1000$ and $S=100$ to directly compare our results with the ones in~\cite{Baldi:2014kfa}. The significant improvement in this case is due to the use of the full ML output in both MLL+KDE and BL methods, while in Ref.~\cite{Baldi:2014kfa} only a fraction of $o(x)$ is used to define a signal enriched region with a working point.

\subsection{SSM $Z^{\prime}$ boson decaying into lepton pairs at the HL-LHC} \label{sec:zprime}

In this section, we analyzed the performance of our method on a simple collider example, namely the search for an SSM $Z^{\prime}$ boson decaying into lepton pairs at the HL-LHC. We generated sample events for signal and background with {\tt MadGraph5\_aMC@NLO 2.6}~\cite{Alwall:2014hca}, the showering and clustering were performed with {\tt PYTHIA 8}~\cite{Sjostrand:2014zea}, and finally, the detector simulation was done with {\tt Delphes 3}~\cite{deFavereau:2013fsa}. For the SM background, we considered the Drell-Yan production of $Z/\gamma{*}$ $\to$ $\ell \ell$, with $\ell$ = $e$, $\mu$, as in~\cite{ATLAS:2018tvr}. As in the previous examples, we trained a {\tt XGBoost} classifier, with 1M events per class, to distinguish $S$ from $B$, for each $Z^{\prime}$ mass value, $m_{Z^{\prime}}$=[2.5, 3.5, 4.5, 5, 5.5, 6.5 , 7.5, 8.5] TeV, and final state (dielectron and dimuon). We use as input parameters the transverse momentum $|p_T|$, the azimuthal angle $\phi$, and the pseudorapidity $\eta$ of the final state leptons in each channel, the kinematic variables that can be extracted directly from the {\tt Delphes 3} output file. Considering the expected background prediction, for each parameter point we scan over $S$ to obtain the expected signal yield upper limit at 95$\%$ C.L., corresponding to the value that gives $Z=1.64$. Finally, we convert this yield to a cross section-upper limit that can be compared with the theoretical prediction.

\begin{figure}
  \centering
   \includegraphics[width=0.48\textwidth]{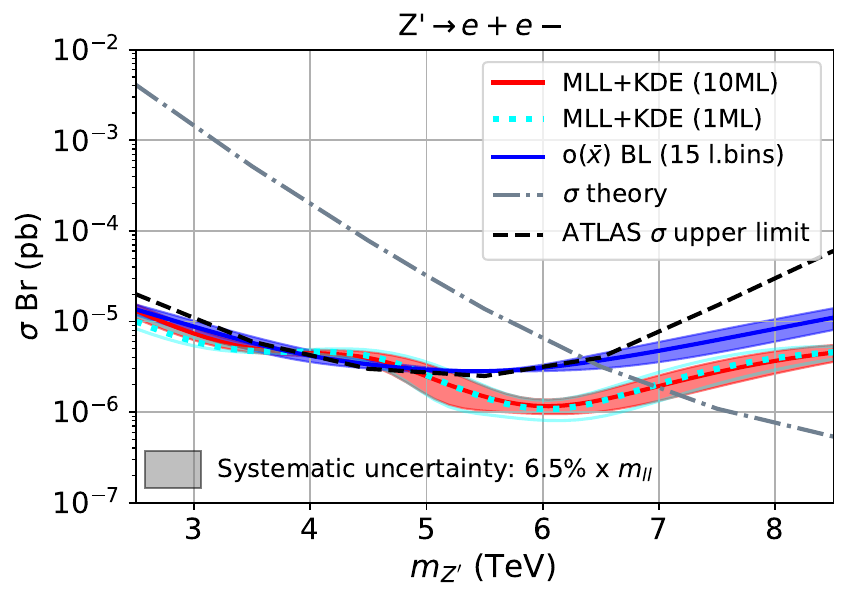}
      \includegraphics[width=0.48\textwidth]{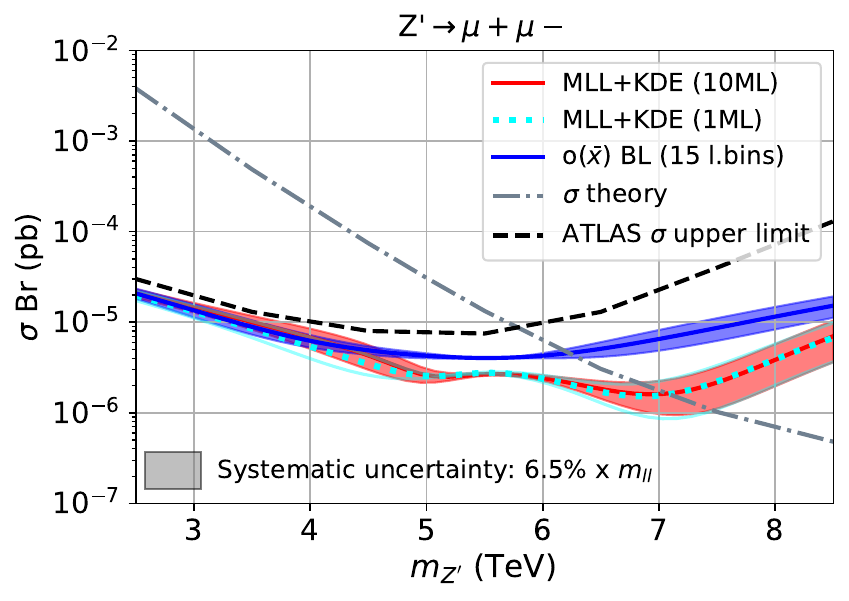}
\caption{Exclusion limits for the $Z'_{SSM}$ with MLL+KDE method (red solid curve corresponds to the averaged variable of 10 ML, and dotted cyan curve to $1$ ML), and with the BL fit of the ML output using 15 linear bins (blue curve). The shaded area in each case includes a naive estimation of the significance uncertainty caused by the mass variation on each point, according to the systematic uncertainty for the invariant mass estimated by ATLAS in~\cite{ATLAS:2018tvr}. \textit{Left panel:} Dielectron channel. \textit{Right panel:} Dimuon channel.
}\label{fig:zprime}
\end{figure}

We are employing the same setup and detector level cuts as in the work presented by the ATLAS Collaboration at $\sqrt{s}=14$ TeV and 3 ab$^{-1}$~\cite{ATLAS:2018tvr}, but we only generated signal and background events with dielectron and dimuon invariant masses above 1.8 TeV, and since we are dealing with a signal-enriched region and not the entire spectrum, the direct comparison with ATLAS projections for 95\% CL exclusion limits is not strictly fair. This may enhance the performance of our classifier, since a $Z^{\prime}$ signal would appear as an excess at high dilepton invariant masses. However, the power of our method can be shown in the left (right) panel of Figure \ref{fig:zprime} for the dielectron (dimuon) channel when compared to the BL fit of the ML classifier output, which is on equal footing with the results for our method since it uses the same ML classifier. Unbinning signal and background posteriors provide more constraining exclusion limits for both final states, and as in the previous examples, there is no significant difference between MLL+KDE using the output of 1ML or the averaged 10 ML.

\section{Conclusions}\label{sec:concl}

The Machine-Learned Likelihoods method can be used to obtain discovery significances and exclusion limits for additive new physics scenarios.
It uses a single classifier and its full one-dimensional output, which allows the estimation of the signal and background PDFs needed for statistical inference.
In this paper, we extend the MLL method to obtain exclusion significances and improve its performance by using the KDE method to extract the corresponding PDFs from the ML output.
We found that the small fluctuations of the machine learning output around the optimal value translate into non-smooth PDFs. We verify that this problem can be handled by averaging the output of several independent machine-learning realizations. But mostly, we show that these small fluctuations do not have a major impact on the final significance.

Although the binning of the classifier output is always possible, irrespective of the dimensionality of the original variables, we verify that computing the PDFs with a non-parametric method such as KDE to avoid the binning, enhances the performance.
By analyzing toy models generated with Gaussian distribution of different dimensions (with and without correlation between signal and background), we showed that MLL with KDE outperforms the BL method (with both linear and non-linear bins) when dealing with high-dimensional data, while for low-dimensional data all the methods converge to the results obtained with the true PDFs. 
Although it is a well-known fact that almost all the benefits of unbinned approaches can be obtained with optimal binning, avoiding such a (usually 
cumbersome) process is one of the main advantages of our work, providing an automatic way of estimating the probability density distributions through the KDE implementation.

Finally, we test the MLL framework in two physical examples.
We found that, as expected, MLL also improves the exclusion-limits results obtained in a realistic $Z'$ analysis as well as in the search for exotic Higgs bosons at the LHC, surpassing the ones computed with the simple BL fit of the ML one-dimensional output.

Last but not least, we would like to remark that this new version of MLL with KDE does not include systematic uncertainties in the likelihood fit, which is necessary for any realistic search. As this is a highly non-trivial issue for unbinned methods, we leave the inclusion of nuisance parameters to the MLL framework for future work. Nevertheless, we also highlight that even though likelihoods without uncertainties can not be used in most experimental setups, it could be useful in specific scenarios where the nuisance parameters can be considered small, and in phenomenological analyses as proofs of concept.

\vspace{2.5mm}
\paragraph{Acknowledgments.}
We thank Xabier Marcano, Manuel Szewc, Alejandro Szynkman, and Hernán Wahlberg for fruitful discussions on the implementation of KDE in the MLL method.
This work is partially supported by the Comunidad Aut\'onoma de Madrid through the ``Atracci\'on de Talento'' program (Modalidad 1) under the grant number 2019-T1/TIC-14019 (EA, RMSS) and through the grant SI2/PBG/2020-00005 (AP, MdlR), and by the Spanish Research Agency (Agencia Estatal de Investigaci\'on) through the grants IFT Centro de Excelencia Severo Ochoa No CEX2020-001007-S (EA, AP, MdlR, RMSS), PID2021-124704NB-I00 (EA, RMSS), and PID2021-125331NB-I00 (AP, MdlR), funded by MCIN/AEI/10.13039/ 501100011033. EA and AP also acknowledge financial support from CONICET and ANPCyT under the projects PICT 2017-2751 and PICT 2018-03682.

\appendix

\section{Automatic tune of binning} \label{sec:appendix}

In the KDE method presented in this work,
we are automatically tuning the bandwidths of the background and signal
distributions. On the other hand, for the histograms, assuming equal-sized bins, there is only one
parameter to tune: the bin width (or equivalently, the number of bins).
It is important to remark that although
we have two samples (background and signal) the formula for significance in Eq.~\eqref{eq:testexclusion2} requires the same binning
for both distributions. In that sense, we have chosen to tune the histogram associated with the
background sample, since it limits the validity of the binned likelihood method in statistical terms.
There are different methods that can be used to tune such parameter (Stone, Freedman Diaconis or FD, Auto, Doane, Scott, Sturges, etc.).
Nevertheless none of these methods is general for every possible ML output as each one is optimized taking into account the functional form of the data (size, shape, etc), which in our case is the  output of our classifier that depends on the specific physical scenario.
For example, the Sturges method is robust for Gaussian data, while FD method is good for large data samples. 

\begin{figure}
  \centering
   \includegraphics[width=0.48\textwidth]{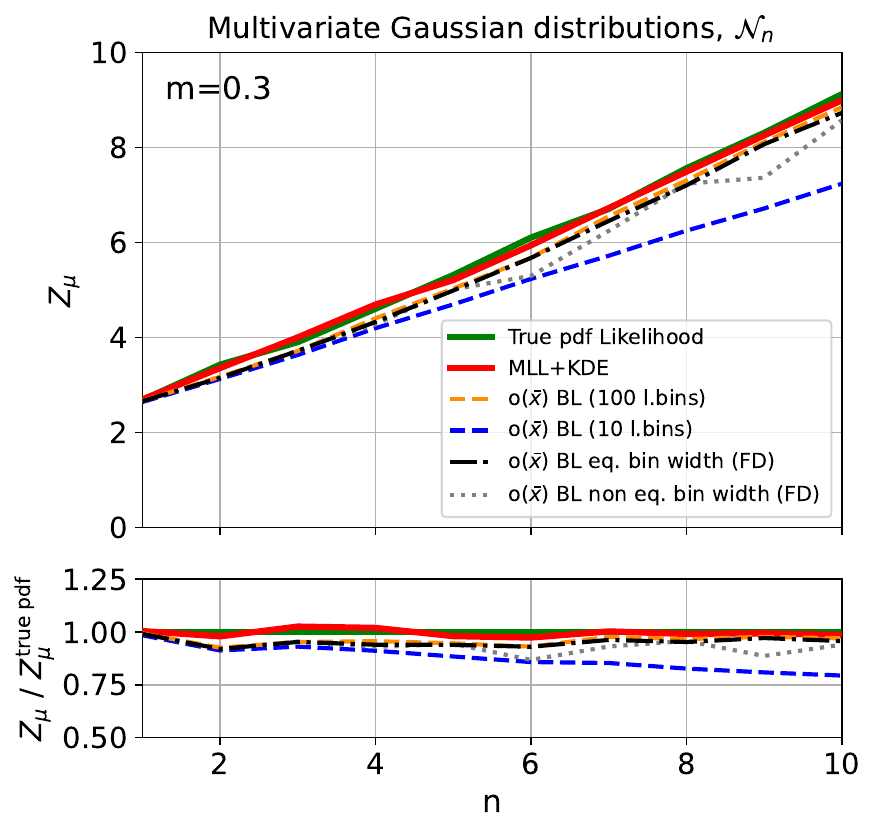} 
      \includegraphics[width=0.48\textwidth]{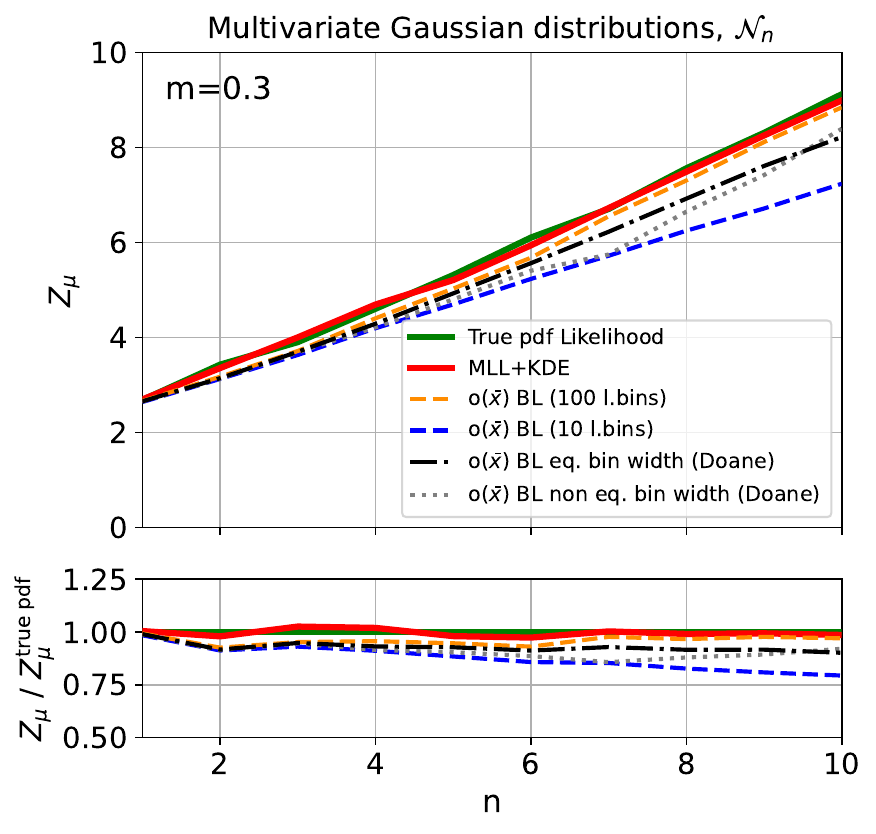} \\
   \includegraphics[width=0.48\textwidth]{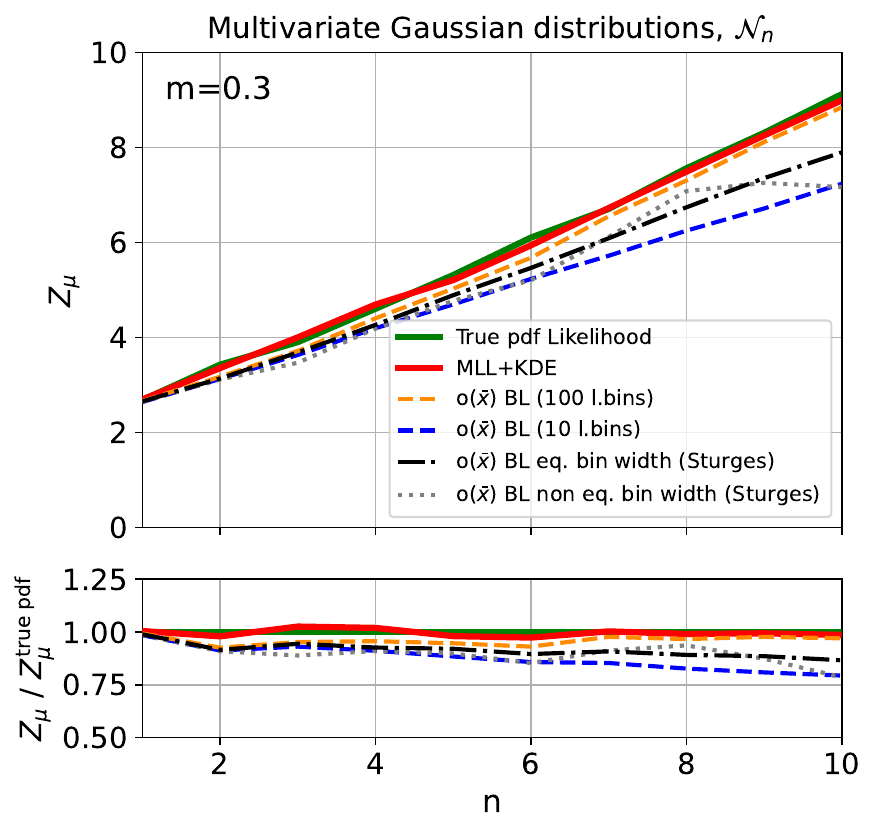}
\caption{Exclusion-limit significance for $\mathcal{N}_{dim}(\boldsymbol m,\boldsymbol\Sigma)$ with $ \boldsymbol m= +0.3 (-0.3)$ for $S$ ($B$) and no correlation, as a function of the $dim$, for fixed $\langle B \rangle=50$k and $\langle S \rangle=500$.  The red solid curve shows the result of implementing the MLL+KDE method, while the green curve shows the results obtained using the true PDFs.
The orange dashed curve represents the results obtained by applying the BL method to the classifier's one-dimensional output for 100 equal-width bins, and the blue dashed curve for 10 equal-width bins. The black dot-dashed curve represents the BL method to the classifier's one-dimensional output for N equal bin width with N determined for each dimension by FD (\textit{top left panel}), Doane (\textit{top right panel}), and Sturges (\textit{bottom panel}), see Table~\ref{table:non-uniform-bins}. The gray dotted curve shows the significance using BL method with N bins, assuming non-equal bin widths.}
\label{dimgaussian-Zbins-binwidths}
\end{figure}

In table \ref{table:non-uniform-bins} we show the optimal number of bins found by three different methods \footnote{We use the python implementation available in the library \href{https://numpy.org/doc/stable/reference/generated/numpy.histogram_bin_edges.html}{\texttt{numpy.histogram$\_$bin$\_$edges}}.} for $\mathcal{N}_{dim}(\boldsymbol m,\boldsymbol\Sigma)$, with $ \boldsymbol m= +0.3 (-0.3)$ for $S$ ($B$), no correlation, and fixed $\langle B \rangle=50$k and $\langle S \rangle=500$, corresponding to the Gaussian example introduced in Section~\ref{sec:pdf}. As stated before, since there is no general method to choose $N$,  we decided to compare three methods whose assumptions fit some of our data set properties: FD due to our large sample, Sturges because at low dimensions the ML output resembles a normal distribution, and Doane to account the skew of the data for high dimensions. 
Additionally, in Figure~\ref{dimgaussian-Zbins-binwidths} we show the significances obtained with these $3$ methods for Gaussian distributions of different dimensions.
It is important to highlight that all these methods assume equal size bin
widths, and hence they show the same tendency already presented in the left
panel of Figure~\ref{dimgaussian-Zbins}, the significance increases
with the number of bins.

\begin{table}[h]
\centering
\begin{tabular}{ c|ccc } 
 \hline
  & \multicolumn{3}{c}{Number of bins (N)} \\ 
  \hline
 Dim & FD & Doane & Sturges \\ 
  \hline
 1 & 74 & 21 & 18 \\ 
 2 & 62 & 23 & 18 \\ 
 3 & 57 & 23 & 18 \\ 
 4 & 54 & 24 & 18 \\ 
 5 & 52 & 24 & 18 \\ 
 6 & 52 & 24 & 18 \\ 
 7 & 52 & 24 & 18 \\ 
 8 & 54 & 25 & 18 \\ 
 9 & 55 & 25 & 18 \\ 
 10 & 57 & 25 & 18 \\ 
 \hline
\end{tabular}
 \caption{Estimation of the optimal number of bins to describe the background data of the example shown in Figure~\ref{dimgaussian-Zbins-binwidths}, using FD, Doane and Sturges methods.}
\label{table:non-uniform-bins}
\end{table}

If we do not assume equal-size bins, we must tune ($N$-1) parameters (i.e. the width of each bin) where $N$ is the number of bins.
Unlike in the KDE method, now we are dealing with a high dimensional space. To perform a full exploration of this space is computationally expensive, therefore we performed a data-driven procedure to cross-validate the selection of the number of bins and each bin size as follows:

\begin{enumerate}
\item We fix the number of bins ($N$) with one of the previously described methods (FD, Doane and Sturges) applied to the background data set.
\item We randomly select the width of each bin.
\item We divide the background data set into 5 k-folds (as done in the KDE optimization). For each bin $d$, we compute the mean number of events per bin, $\mu_d = \frac{1}{4} \sum_{k=1}^{4} N_d^{(k)}$, where $N_d^{(k)}$ is the number of events in each bin for the sample $k$. Notice that we used 4 k-folds.
\item For the last k-fold, $k=5$, assuming a Poissonian distribution for each bin and the Stirling approximation for $\text{log}(N_{d}!)$, we calculate
\begin{equation*}
    q_{\text{poiss}} = -\text{log} \mathcal{L} = \sum_d^{N_{bin}} \, ( \, \mu_d - N_d^{(k=5)} \text{log}(\mu_d) + N_d^{(k=5)} \text{log}(N_d^{(k=5)}) - N_d^{(k=5)} \, ).
\end{equation*}
\item We repeat steps 2. to 4. 5000 times.
\item Finally, we select the binning that provides the minimum value of $q_{\text{poiss}}$ from among the 5000 iterations and compute the significance using the signal and an independent background sample.
\end{enumerate}
This procedure provides a good trade-off between optimization and computational cost. In Figure~\ref{dimgaussian-Zbins-binwidths} we show the obtained exclusion significance ($Z$).
For each dimension, the significance found optimizing the bin widths is similar to the result assuming equal-sized bins (for the same number of bins). 
It is worth to remark that the MLL-KDE method consistently outperforms the binned multivariate analysis in terms of significance.
For completeness, we have also check that similar conclusions can be drawn for alternative likelihood assumptions (for example, assuming a Gaussian distribution for each bin).


\bibliographystyle{JHEP} 
\bibliography{biblio}

\end{document}